# MODELING CASSAVA YIELD: A RESPONSE SURFACE

A.O. Bello
Federal University of Technology, Nigeria

### ABSTRACT

*This paper reports on application of theory of experimental design using graphical techniques in R programming language and application of nonlinear bootstrap regression method to demonstrate the invariant property of parameter estimates of the Inverse polynomial Model IPM in a nonlinear surface.*

## 1. INTRODUCTION

Yield has always being a major yard stick for measuring successful agricultural method and economic indices.

There has being increased demand for export of cassava to China, US and other developed nations of the world whose production of currency papers, starch related products, recently fast increasing bio-fuel energy and others, largely depends on Africa's production of cassava. The demand for Cassava has globally increased and it overshoot supply, the occurrence of drop in yield has put a lot of pressure on production of Cassava and the present increase in cultivation is not enough to curb demand, according to Food and Agriculture Organization of the United Nations database FAOSTAT (2009.

Achieving the production efficiency of cassava is the problem that this research tends to employ statistical tools to approach, focusing on the area of modelling Cassava yield plantation and highlighting the respective point of intercept and gradient levels of fertilizer and crop spacing with the variety of cassava that is capable of delivering the cassava production efficiency. The soil, cultural practices, and socio-economic factors also play a key role in cassava production. In many tropical countries, soil fertility is usually poor and requires fertilization to achieve a high level of cassava yield M. Muengula-Manyi1, et al., 2012 . Omorusi2011 on the other hand reported that the incidence of CMD infection that affects cassava yield was depressed by high dosage of NPK (100 kg ha-1 when compared to plants in soil with low NPK (10 kg ha -1.

### 1.2 The World Bio-Fuel Target: The Pressing Pressure for High Cassava Yield.

According to Rishi Sidhu2011, cassava is the newest addition to the bio-fuel line-up, joining other bio-chemical substances, and has doubled in price because of its new role, quoting New

York Times. To find a new source of energy, Chinese scientists perfected a process to make fuel from cassava. The European Union has said that 10 percent of transportation fuel must come from renewable sources like bio-fuel or wind power by 2020, and the U.S. requires bio-fuel use to reach 36 billion gallons annually by 2022. These targets, according to Timothy D. Searchinger(2013) a research scholar at Princeton University. Failure to have quick increase in yield would not only affect the bio-fuel target but also amount to cassava food and its by-products shortage.

## 2. METHODOLOGY

RSM involves several experiments, and using the results of one experiment to provide direction for what to do next. This next action could be to focus the experiment around a different set of conditions, or to collect more data in the current experimental region in order to fit a higher-order model or confirm what we seem to have found (Russell V. Lenth, 2012).

$$y = f(x)\beta + \varepsilon \qquad (1)$$

Where $x = (x_1, x_2, ..., x_k)$, $\beta$s is the coefficient estimates, $\varepsilon$ (error term) ~ N (0; 1) and independent. $f$ is the function that describes the form in which the response and the input variable are related, its mathematical form is not known in practice. It is often approximated within the experimental region. The purpose of considering a model such as Eq. 1 is to establish a relationship, albeit approximations, that can be used to predict response values for given settings of the control variables, determine significance of the factors. The totality of the predictors' settings is denoted by D, considering Eq. 1, in matrix form;

$$y = X\beta + \varepsilon \qquad (2)$$

Ordinary least-squares estimator of $\beta$ is $\hat{\beta} = (X'X)^{-1}X'Y$

Y, estimates variance $Var(\hat{\beta}) = \delta^2 (X'X)^{-1}$ and the predicted response given as $\hat{y}(x) = f, \hat{\beta}(x) \in \Re$ with variance

$$Var[\hat{y}(x)] = \delta^2 (X'X)^{-1} f(x) \qquad (3)$$

The proper choice of design is very important in any response surface investigation. This is true because the quality of prediction, as measured by the size of the prediction variance, depends on the design matrix D as can be seen from formula $Var(\hat{\beta})$. It is therefore imperative that the prediction variance in Eq. 3 be as small as possible provided that the postulated model in Eq. 1 does not suffer from lack of fit. [See Andre I Khuri et al (2010) and Khuri, AI, Cornell (1996) for details].

### 2.1 Orthogonality

A design D is said to be orthogonal if the matrix XX is diagonal, where X is the model matrix in Eq. 2. The elements of $\hat{\beta}$ will be uncorrelated because the off diagonal elements of $Var(\hat{\beta})$ will be zero.

## 2.2 Rotatability:

A design D is said to be rotatable if the prediction variance in Eq. 3 is constant at all points that are equidistant from the design centre. The prediction variance remains unchanged under any rotation of the coordinate axes.

*C. Uniform Precision:*

A rotatable design is said to have the additional uniform precision property if Var $[\hat{y}(x)]$ at the origin is equal to its value at a distance of one from the origin. This helps in producing some stability in the prediction variance in the vicinity of the design centre

Fig. 1.
$$D = \begin{pmatrix} x_{11} & . & . & . & x_{1k} \\ . & . & & & . \\ . & & . & & . \\ . & & & . & . \\ x_{n1} & . & . & . & x_{nk} \end{pmatrix}$$

## 2.2 The Choice Of Design and Model:

The problem of establishing cassava production efficiency in a world of inadequate resources, where resources such as man-power, seedling, fertilizers, time etc cannot be risked to test run all possible yield in a real life planting venture emphasis on tropical regions of Africa, where other agricultural factors are asymptotically the same. Parsad, R. et al (2000), noted that the cost of constructing a response surface affects the choice of design of experiment and this can have large influence on accuracy of the approximation. These call for caution in making choice of design and model.

In the quest of modelling cassava yield effectively, a good choice of design and model (methodology) is vital. The above challenges suggest the need for a methodology that has the ability to develop a smaller design that can lead to a model that will adequately suggest all possible yields per planting input factors. A design that will help to reduce the number of experimental runs which could be costly or unattainable. The ability to model all possible yields per planting input will help to estimate the quantity of harvest that will eventually be due for exports and local consumption. Also this will ensure good planning for each farming and economic year.

## 2.3 The Design

The first-order design proposed for fitting first-degree model-[factorial Design]. The second-order design proposed for fitting second-degree model-[The Central Composite design (CCD)] - Orthogonality, Rotatability and Uniform Precision are desired properties in any design. CCD consist the full factorial runs (or fractional factorial of resolution V) with nf runs, also called "cube point", (ii) The axial or star runs consisting $n_a = 2k$ points on the axis of each factor at a distance from the center of the design and, (iii) The center runs $n_0$ (0, 0..., 0) without replication. The total number of runs $N = n_f + n_a + n_0$ (See Klaus Hinkelmann, Design and Analysis of Experiments Volume 1 Second Edition for details).

$$b = \sum_{i=1}^{N} x_i^2 = \sum_{i=1}^{N} x_j^2, c = \sum_{i=1}^{N} x_i^2 x_j^2 \text{ and } \qquad c + d = \sum_{i=1}^{N} x_i^4 = \sum_{i=1}^{N} x_j^4. \qquad (5)$$

For appropriate distance from the center $\alpha = (nf)^{1/4}$

$$b^2 = Nc \Leftrightarrow [nf + 2\alpha^2]^2 = [nf + n_a + n_0]n_f$$

and

d=2c  (6)

## 2.4 The Inverse Polynomial Model (IPM) A Class of Nonlinear Model

The general forms of Inverse Polynomials of (Nelder, 1966; Nduka, 1994; Holger Dette, 2007) are family of non-linear function.

$$\frac{x_1, x_2, ..., x_k}{y} = polynomial \text{ in } x_1, x_2, ..., x_k \quad \text{(Nelder1966)}$$

$$\prod_{i=1}^{k} \frac{x_i}{q} = \prod_{i=1}^{k} (\beta_0 + \beta_{1i} x_i + \beta_{2i} x_i^2 + ... + \beta_{p-1i} x_i^{p-1}) \quad (7)$$

Where, k = number of factors  p = number of levels of factor i.   q = the yield per unit area.

Intrinsically Non-linear Model; Non-linearity in parameters and when linearity cannot be achieved through transformation

A regression model is called non-linear, if the derivatives of the model with respect to the model parameters depend on one or more parameters. Method of Gauss-Newton is more adequate in estimation of inverse polynomial model; it is preferable because of its faster convergence in inverse polynomials, also the boundedness, invariant of parameter estimates, distribution free, asymmetry, and quick location of optimum properties favoured the choice of Inverse Polynomial model. [See Nduka, E. C. 1994 and 1997].

*First Order Model;

$$\prod_{i=1}^{k} \frac{x_i}{q} = (\beta_{01} + \beta_{i1} x_1)(\beta_{02} + \beta_{i2} x_2) \quad \text{also, let q = y}$$

$$\frac{x_1 x_2}{q} = \beta_{01}\beta_{02} + \beta_{01}\beta_{12} x_2 + \beta_{11}\beta_{02} x_1 + \beta_{11}\beta_{12} x_1 x_2$$

$$y^{-1} = \beta_{11} + \beta_{01} x_1^{-1} + + \beta_{10} x_2^{-1} + + \beta_{00} (x_1 x_2)^{-1}$$

*Second Order Form Quadratic Surface (k=2, p=4)

$$\frac{x_1 x_2}{q} = (\beta_{01} + \beta_{11} x_1 + \beta_{21} x_1^2)(\beta_{02} + \beta_{12} x_2 + \beta_{22} x_2^2)(\beta_{03} + \beta_{13} x_1 + \beta_{23} x_1^2)(\beta_{04} + \beta_{14} x_2 + \beta_{24} x_2^2)$$

above is rewritten considering only up to the 2nd order as ;

$$y^{-1} = \beta_{11} + \beta_{01} x_1^{-1} + \beta_{10} x_2^{-1} + \beta_{02} x_2 x_1^{-1} + \beta_{20} x_1 x_2^{-1} + \beta_{00} (x_1 x_2)^{-1}$$
(10)

## 3. PARAMETER ESTIMATION

*Nonlinear Estimation*

The normal equation is extremely difficult to solve, if it have multiple solution to multiple stationary values of function $S(\hat{\theta})$. It is said to have no closed form solution. Thereby the procedure of iterative method must be approached in order to man over the mathematical intractable problem.

### 3.1 Inverse Polynomial Nonlinear Estimation

Considering the parameter $\theta$, the estimate ($\hat{\theta}$) of θ is obtained by differentiating equation S ($\hat{\theta}$) and equated to zero, giving rise to J-normal equations and solve for the θs. Considering determistic component of our model in Eq. 1. The sum of square error for non linear model is defined as

$$S(\theta) = \sum_{i=1}^{n}[y_i - f(x_i, \theta)]^2 \qquad (11)$$

When the J-normal equation becomes mathematical intractable, an iterative procedure is required.

$$S(\theta) = \sum_{i=1}^{n}[y_i - f(x_i, \theta_j)]^2$$

Let $\beta_j = \theta_j$

$$S(\theta) = \sum_{i=1}^{n}[y_i - (\theta_{00}(x_1 x_2)^{-1} + \theta_{01} x_2^1 + \theta_{11})^{-1}]^2 \mid \frac{\partial f(x_i, \theta)}{\partial \theta_j} \mid_{(\theta_j - \theta_j^0)}$$

Combining $\dfrac{\partial S(\theta)}{\partial \theta_{00}}, \dfrac{\partial S(\theta)}{\partial \theta_{01}}, \dfrac{\partial S(\theta)}{\partial \theta_{10}}$ and $\dfrac{\partial S(\theta)}{\partial \theta_{11}}$ we the normal equation;

$$\sum_{i=1}^{n}[y_i - f(x_i \theta)][\theta_{00}(x_1 x_2)^{-1} + \theta_{01} x_1^{-1} + \theta_{10} x_2^{-1} + \theta_{11})^{-2}](x_1 x_2)^{-1} = 0$$

$$\sum_{i=1}^{n}[y_i - f(x_i, \theta)][\theta_{00}(x_1 x_2)^{-1} + \theta_{01} x_1^{-1} + \theta_{10} x_2^{-1} + \theta_{11})^{-2}]x_1^{-1} = 0$$

$$\sum_{i=1}^{n}[y_i - f(x_i, \theta)][\theta_{00}(x_1 x_2)^{-1} + \theta_{01} x_1^{-1} + \theta_{10} x_2^{-1} + \theta_{11})^{-2}]x_2^{-1} = 0$$

$$\frac{\partial S(\theta)}{\partial \theta_{11}} = 0$$

The normal equation has no close form solution.

### 3.2 Gauss Newton Numerical Method

Consider the deterministic component of equation (1), by the Taylor first-order approximation;

$$f(x, \theta) = f(x, \theta^0) + \frac{\delta f(x, \theta)}{\delta \theta}|(\theta - \theta^0) \qquad (12)$$

by substitution,
$y = f(x, \theta) + \varepsilon$

$$y = f(x, \theta^0) + \frac{\delta f(x,\theta)}{\delta \theta}|(\theta - \theta^0) + \varepsilon$$

For the j parameters and i$^{th}$ observation, we have;

$$y - f_i^0 = \sum_{j=0}^{j-1} \frac{\delta f(x_i,\theta^0)}{\delta \theta_j}|(\theta - \theta^0) + \varepsilon \qquad (13)$$

Note that $f_i^0 = f(x_1, x_2, ..., x_k, \theta^0_0, \theta^0_1, ..., \theta^0_{j-1})$

Let $\beta_j^0 = (\theta_j - \theta_j^0)$ and $Z_{ij} = \frac{\delta f(x_i,\theta^0)}{\delta \theta_j}|(\theta_j - \theta_j^0)$

Eq. 13 can be written as

$$y_i - f_i^o = \sum_{i=i}^{k} \beta_j Z_{ji} + \varepsilon_i$$

The above is also the same in matrix form below

$$Y - f^0 = Z^0 \beta + \varepsilon \qquad (14)$$

Column vector $Y - f^0_{(k \times 1)}$, $Z^0_{(j \times k)}$ is determinant matrix, $\beta_{(j \times 1)}$ is the coefficient matrix and $\varepsilon$ is the error term(kx1). The least square estimator is of form

$$\hat{\beta} = (Z^{O'}Z^O)Z^O(Y - f^0) \qquad (15)$$

minizing the sum of square $S^0(\theta) = \sum_{i=1}^{n}[y_i - f_i^o - \sum \beta_j Z_{ij}]^2$

The initial value for $\theta$ $tn$ $\beta_j^0$ and $Z_{ij}$ is supplied either through theoretical knowledge of subject, intelligent guess, self-starter function in R or grid search function in SAS. A good choice of starting value will ensure rapid convergence if not there may not be any convergence in the series. (See Schabenberger Oliver,"Nonlinear Regression in SAS").
Condition for Convergence

$$|\frac{(\theta_{i(j+1)} - \theta_{ij})}{\theta_{ij}}| < \delta, = 0.000001 \text{ or } 1.0 \times 10^{-6} \qquad i=1,2,...,n$$

### 3.3 Gauss Newton Nonlinear Least Square Estimation of IPM

Starting value for θj s i.e ($\theta_{00}, \theta_{01}, \theta_{10}$ and $\theta_{11}$) is supplied for 1st iteration using equation[19], note and $\psi = \theta_{00}(x_1 x_2)^{-1} + \theta_{01} x_1^{-1} + \theta_{10} x_2^{-1} + \theta_{11})^{-2}$

### 3.4 The general Bootstrap Algorithm.

1. Generate a sample x$^{*'}$ of size k from distribution F with replacement.
2. Compute $\theta_j$ for this bootstrap sample.
3. Repeat steps 1 and 2, B time.
By this procedure we end up with bootstrap values

$$\theta_{bj}^* = [\theta_{b0}^*, \theta_{b1}^*, ..., \theta_{bk}^*]' \quad j=1,...,B$$

Afterward, we can use these bootstrap values to calculate all the quantities of interest. Generally, 500 or 1000 samples will be sufficient [See Dimitris Karlis (2004) for details].

## 4. NUMERICAL ANALYSIS.

The data considered for this work is an experimental data of two factorial design, graded into four levels, having crop spacing (meters) and inorganic fertilizer application per crop stand (decilitres). Cassava varieties TMS 30572 were exposed to the factors (crop spacing and fertilizer). Plant and soil analysis were also carried out to ensure the required agricultural standards. The responses generated were measured in yield (kilograms) and dimension (centimetres) as dual responses Yield and Dim respectively.

### 4.1 Bootstrap Re-sampling Method

The term bootstrap, coined by B. Efron (1986, and 1997), refers to using the sample to learn about the sampling distribution of a statistic without reference to external assumptions while it is a method for improving estimators, it is also well known as a method for estimating standard errors, bias and constructing confidence intervals for parameters. It substitute considerate amount of computation in place of theoretical analysis (B.Efron and Tibshirani,1986).

TABLE I.  R-NLS( ) ITERATION RESULT

| IPM | $\beta_S$ | iteration | SSE | Convergence Tolerance | Pr>F |
|---|---|---|---|---|---|
| Yield | 6 | 10 | 2.25 | 7.47e-07 | 0.001 |
| Dim | 6 | 12 | 349.2 | 5.62e-07 | 0.001 |

### 4.2 Model Adequacy

The Shapiro-Wilk normality test is conducted for cassavaYield and cassavaDim models respectively; standardRes results W = 0:9386, W = 0:9767 respective closeness of standardRes values to '1' indicated the assumption of normally distributed errors is not violated.

TABLE II.  BOOTSTRAP RESULT WITH YIELD

| Parameter | Bootstrap Sample | | Observed Sample | |
|---|---|---|---|---|
| | Estimates | C.I (2.5% 97.5%) | Estimates | C.I (2.5% 97.5%) |
| $\beta_{11}$ | 0.356039089 | 0.2  0.6 | 0.349449160 | 0.1  0.62 |
| $\beta_{01}$ | -0.00922816 | -0.1  0.12 | -0.00844514 | -0.09  0.07 |
| $\beta_{10}$ | -0.22006195 | -0.54  0.09 | -0.21609962 | -0.66  0.21 |
| $\beta_{00}$ | 0.011454663 | -0.05  0.078 | 0.011312689 | -0.07  0.099 |
| $\beta_{20}$ | 0.202227135 | -0.242  0.617 | 0.197842082 | -0.386  0.793 |

First and second CassavaYield Model;

$$1^{st} = [0.356 - 0.0092x_1^{-1} - 0.2201x_2^{-1} + 0.0115(x_1x_2)^{-1}]^{-1}$$
$$2^{nd} = [0.356 - 0.0092x_1^{-1} - 0.2201x_2^{-1} + 0.0115(x_1x_2)^{-1} + 0.2022x_1x_2^{-1}]^{-1}$$

## 5. DISCUSSION OF RESULTS

The various residual plots and test has giving a basis to submit that the inverse polynomial model is appropriate for fitting cassava yield (kg) and cassava dimension (cm).

It was also observed that the residuals does not sum up to zero as of the cases of the linear surfaces. This is expected of a non-linear result. The results of bootstrap and the observed data show closer agreement, which produce a negligible bias value. The bootstrap C.I. and original sample C.I also agree quite well demonstrating the precision of the model coefficient estimates. The strong significance of crop spacing in the model of cassava plantation has shown that crowding of plants or mixed planting should be discouraged. The obtained non-linear estimates were reintroduced as the starting values in model and the same estimates value were obtained with their sum of squares error unchanged and zero iteration number, indicating a global minimum estimates were obtained.

## 5.1 CONCLUSION

It is acceptable to say the cassavaYield model is most adequate, robust and best to fit cassava yield and good crop spacing affects cassava yield in a direct proportion.

## ACKNOWLEDGMENT

The author will like to express his appreciation to Prof. T.A. Bamiduro, University of Ibadan, Nigeria for his helpful comments and suggestions.

## REFERENCES

[1] Andre I Khuri and Siuli Mukhopodhgay Advance Review Response Surface methodology Volume 2 March/April 2010;JohnWiley and sons Inc.
[2] B.Efron and Tibshirani(1986) Bootstrap Methods for standard errors, confidence interval and other measure of Statistical Accuracy; Statistical Science 1986 Vol No. 1, pg 54-77
[3] B.Efron and Tibshirani(1997) Improvements of cross-Validation; Bootstrap Method;Journal of the American Statistics Association 92(438)pg 548-560
[4] Christian Ritz and Jens Carl S.(2008) Non-linear Regression with R Springer Science+Business Media LLC
[5] Dimitris Karlis (April 2004)Department of Statistics Athens University of Economics Lefkada, An introduction to Bootstrap Methods 17th Conference of Greek Statistical Society http://stat-athens.aueb.gr/~karlis/lefkada/boot.pdf
[6] FAOSTAT (2009);http://faostat.fao.org Food and agriculture Organization (FAO) of the United Nations, Rome, Italy.
[7] Gary W. Oehlert (University of Minnesota) A First Course in Design and Analysis of Experiments 2000 Edition.
[8] Holger Dette and Christine Kiss (2007) Optimal Experimental Designs for Inverse Quadratic Regression Models; Ruhr-University Bochum
[9] J. A. Nelder(1966) Weight Regression Quantal Response Data, and Inverse Polynomials National vegetable Research Station, Wellesbourne, Warwick, England.

[10] J. A. Nelder(1968) Inverse Polynomials, A useful Group of Multi-Factor Response Functions; The National vegetable Research Station, Wellesbourne, Warwick, England
[11] Jennifer Abraham Cassava: A Plethora Of Opportunities -Copyright © 2013 Punch Nigeria Limited • January 10, 2013
[12] Klaus Hinkelmann ,(State University Department of Statistics  Blacksburg, VA)and Oscar Kempthorne (Iowa State University  Department of Statistics  Ames, IA) - Design and Analysis  of Experiments  Volume 1 Introduction to Experimental Design  Second Edition
[13] Khuri, AI, Cornell, JA. Response Surfaces. 2nd edition.New York: Dekker; 1996.
[14] M. Muengula-Manyi1, et al., 2012 Application of response surface methodology to the optimization of amylase production by Aspergillus oryzae MTCC 1847 American Journal Of Experimental Agriculture 2(3): 336-350, 2012 Science Domain International
[15] Nduka, E.C. and T.A. Bamiduro, 1997. A new generalized inverse polynomial for response surface designs. J. Nig. Stat. Assoc., 11: 41-48.
[16] Nduka, E. C. (1994) Inverse polynomials in the Exploration of Response Surfaces, A Phd Thesis,U. I
[17] Omorusi, V.I., Ayanru. D.K.G. (2011). Effect of NPK fertilizer on diseases, pests, and mycorrhizal symbiosis in Cassava. Int. J. Agric. Biol., 13, 391-395.
[18] Parsad, R., Srivastava, R. and  Batra, P.K. (2004). Designs for Fitting Response Surfaces in Agricultural Experiments. IASRI, Publication
[19] Rajender Parsad And P.K. Batra(2000) Response Surface Design I.A.S.R.I., Library Avenue, New Delhi-110 012
[20] Rishi Sidhu -Cassava, The Latest Biofuel-April12th,2011http://www.Foreignpolicyblogs.Com/2011/04/12/Cassava.
[21] Russell V. Lenth, (The University of Iowa) Response-Surface Methods in R, and illustration Using rsm Updated to version 2.00, 5 December 2012
[22] Russell V. Lenth, (The University of Iowa)Surface Plots in the rsm Package Updated to version 2.00, 3 December 2012
[23] Schabenberger Oliver,"Nonlinear Regression in SAS"  University of California.http: www.ats.ucla.edu/stat/mult_pkg/faq/general/citingats.htm
[24] Timothy D. Searchinger(2013) Bioenergy writings www.scholar.princeton.edu/tsearch/page
[25] Whitcomb and Anderson Excerpted from manuscript for Chapter 10 of RSM Simplified, See $www.statease.com/rsm_simplified.html$

**Appendix**

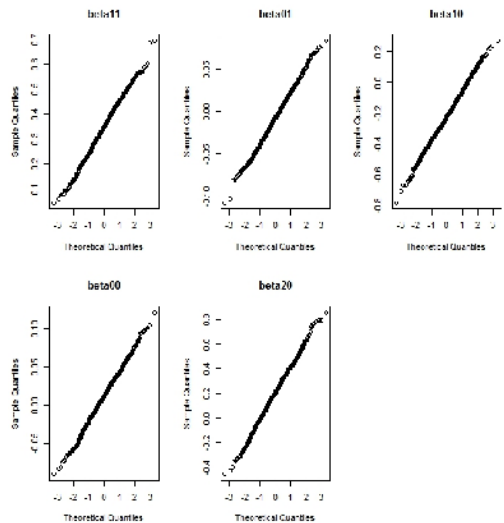

fig1

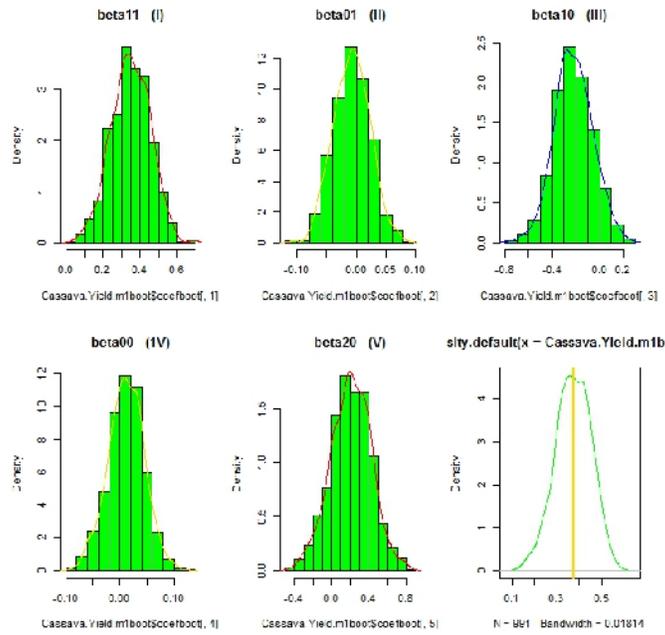

fig ii

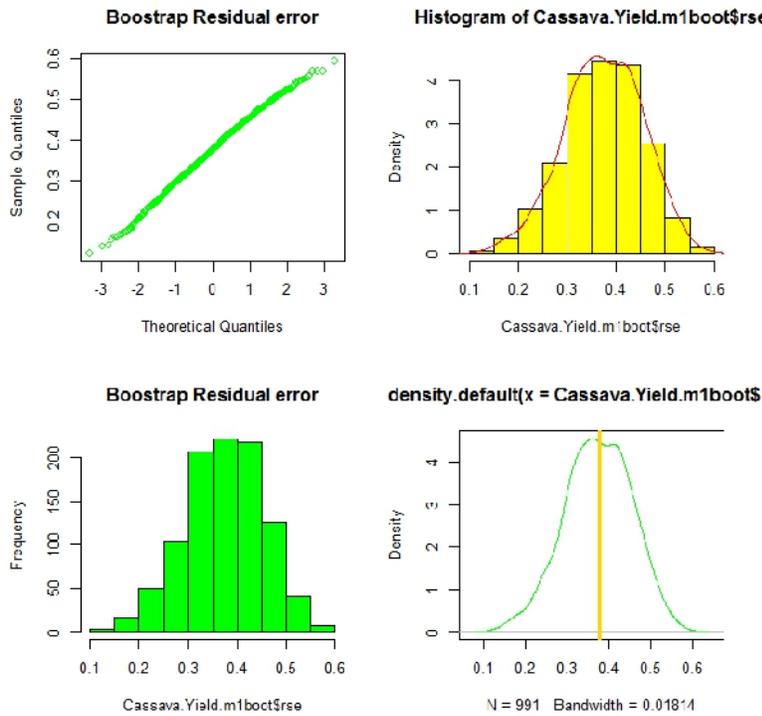

Fig III

## Author

The Author is trained in the discipline of statistics from the cradle of National diploma and Master degree in Statistics from University of Ibadan, Nigeria. He is still under training, His fleer for Design of an Experiment and Nonlinearity studies produced this work. His fleer is also extended to Envirometric research.

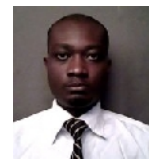